\documentclass[floatfix,aps,prl,twocolumn,superscriptaddress,showpacs]{revtex4-1}
\pdfoutput=1
\usepackage[normalem]{ulem}
\usepackage{float}
\usepackage[usenames]{color}
\usepackage{graphicx}
\usepackage{ulem}
\usepackage{epstopdf}
\usepackage{amsfonts}
\usepackage{amsmath}
\begin{document}

\title{Dynamical behavior of disordered spring networks}

\author{M. G. Yucht}
\affiliation{Princeton University, Princeton, NJ 08544, USA}
\author{M. Sheinman}
\affiliation{Department of Physics and Astronomy, Vrije Universiteit, Amsterdam, The Netherlands}
\author{C. P. Broedersz}
\email{cbroeder@princeton.edu}\affiliation{Lewis-Sigler Institute for Integrative Genomics and the Department of Physics, Princeton University, Princeton, NJ 08544, USA}

\date{\today}

\begin{abstract}
  We study the dynamical rheology of spring networks with a percolation model constructed by bond dilution in a two-dimensional triangular lattice. Hydrodynamic interactions are implemented by a Stokesian viscous coupling between the network nodes and a uniformly deforming liquid. Our simulations show that in a critical connectivity regime, these systems display weak power law rheology in which the complex shear modulus scales with frequency as $G^*\sim (i \omega)^{\Delta}$ where $\Delta = 0.41$, in discord with a mean field prediction of $\Delta = 1/2$.  The weak power law rheology in the critical regime can be understood from a simple scaling relation between the macroscopic rheology and the nonaffine strain fluctuations, which diverge with vanishing frequency for isostatic networks. We expand on a dynamic effective medium theory, showing that it quantitatively describes the rheology of a diluted triangular lattice  far from isostaticity; although the EMT correctly predicts the scaling form for the rheology of near-isostatic networks, there remains a quantitative disparity due to the mean-field nature of the EMT. Surprisingly, by connecting this critical scaling of the rheology with that of the strain fluctuations, we find that the dynamical behavior of disordered spring networks is fully determined by the critical exponents that govern the behavior of elastic network in the absence of viscous interactions.
\end{abstract}

\maketitle

\noindent
Disordered mechanical networks are used to model a variety of 
systems 
including network glasses~\cite{Thorpe1983, Phillips1981, He1985, 
Schwartz1985,Feng1984,Arbabi1993, Feng1985}, jammed packings~\cite{LiuNag1998, Ohern2003,Wyart2008a, 
Ellenbroek2008} and semiflexible biopolymer networks~\cite{Head2003,Wilhelm2003,Heussinger2006, 
Das2007,  Broedersz2011,Mao2011}. Though the specific responses of different classes of networks may vary, the general character of the mechanical behavior depends on the  network's connectivity. Above the so-called isostatic connectivity, networks are 
mechanically stable and resist static shear 
deformations~\cite{Maxwell1864}. The lack of mechanical rigidty below this connectivity is due to zero-energy, floppy deformation modes; however, the mechanical response can be stabilized by additional 
weak interactions~\cite{Wyart2008a,Schwartz1985,He1985, Broedersz2011, Das2012} or internal 
stresses~\cite{Alexander,Sheinman2012}. 
Recently, focus has shifted towards  the dynamic behavior of networks at the verge of mechanical stability with additional viscous interactions \cite{Heussinger2009, Tighe2011, Tighe2012,Tighe2012b,Andreotti2012, Lerner2012a, Lerner2012b, Wyart2010,During2013}.
Marginally stable spring networks in jammed configurations exhibit a rich dynamic mechanical response in the presence of damping 
forces~\cite{Tighe2011,Tighe2012,Tighe2012b}; in the vicinity of the isostatic connectivity, the dynamic 
shear modulus was found numerically to scale with frequency as 
$\sqrt{\omega}$, in accord with a mean field prediction. In contrast, bond-diluted
lattice-based networks---a major class of mechanical 
systems distinguishable from jammed packings~\cite{Ellenbroek2008}---do not exhibit mean-field behavior~\cite{Feng1984,Schwartz1985,Arbabi1993, Broedersz2011}. This class of 
models has provided insight in both the linear and nonlinear elastic response 
of disordered spring and fiber networks.
However, little is understood about the dynamical response of such systems in the presence of viscous coupling between the network and a liquid.

\begin{figure}[h!]
  \begin{center}
    \includegraphics[width=\columnwidth]{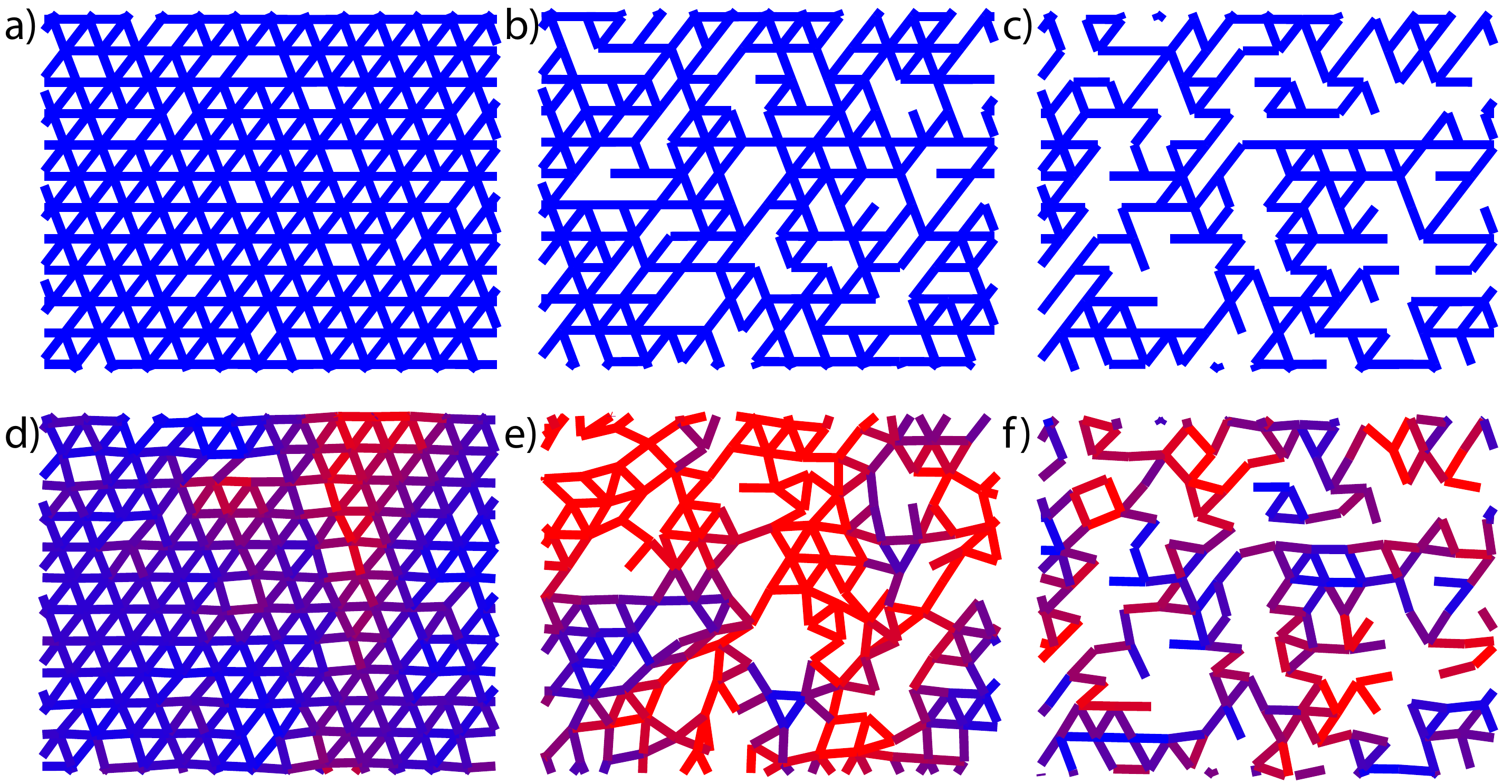}
    \vspace{-0.1in}
    \caption{Exemplary samples of networks undergoing dynamic deformation,
        where the coloring of the plots is a qualitative representation of 
        the nonaffinity of the two nodes connected by each bond (redder is 
        more nonaffine). 
        Networks with $p \gg p_c$ (ad), $p \approx p_c$ (be), and $p \ll p_c$ 
        (cf) are illustrated; the top row depicts high-frequency oscillation, $\omega \gg \omega^*$,
        and the bottom row depicts low-frequency oscillation, $\omega \ll \omega^*$.
        In the high frequency limit, all networks deform affinely because the 
        spring network couples strongly with the affine deformation of the viscous 
        network (abc). In the low frequency limit, the spring network deformation is nonaffine, and the 
        degree of nonaffinity increases as $p$ approaches the isostatic bond 
        probabilty $p_c$ (def). 
    }
    \label{fig:networks}
  \end{center}
\end{figure}

Here we study the linear rheological behavior of percolating 
disordered spring networks immersed in a Newtonian liquid.  
Disordered spring networks are constructed on a triangular lattice in 2D, diluting bonds to control the network's 
connectivity, ranging from well below to well above the isostatic connectivity. 
Analysis of the simulated complex shear modulus of such networks reveals three dynamical 
regimes. For connectivities well above isostaticity, the networks behave as 
solids with weak viscous coupling to the fluid.  The deformation of these 
networks is nonaffine at low frequencies, but becomes increasingly affine at 
high frequencies, as illustrated in Fig.~\ref{fig:networks}.  The second regime is in the vicinity  of the 
isostatic connectivity, where we observe weak power law rheology 
over a frequency range extending to zero-frequency at the isostatic connectivity; this critical slowing down indicates diverging relaxation timescales. In this critical 
regime the complex shear modulus scales with frequency as $G^*\sim(i \omega 
)^\Delta$, where $\Delta \approx 0.41$. In the third regime, with 
connectivities well below isostaticity, the networks behave as Maxwell 
fluids, crossing over from a fluid-like to a solid-like response at 
a connectivity-independent characteristic frequency. Our results are qualitatively consistent with results by Tighe~\cite{Tighe2011,Tighe2012,Tighe2012b} on disordered spring networks based on jammed configurations, although in those 
systems the dynamic exponent $\Delta =1/2$ was found at near-isostatic connectivities.  
In addition, we construct a framework that builds on a dynamic effective medium theory (EMT)~\cite{Wyart2010,During2013} for the rheological response of bond-diluted lattice-based networks and compare this directly with numerical results over a broad range of network connectivities. This dynamic EMT serves as a framework that can be expanded to bond-bending and fiber networks~\cite{Arbabi1993,He1985,Das2007,Broedersz2011,Mao2011,Das2012}. Our EMT calculation for a bond-diluted triangular lattice, taken together with scaling arguments, indicates that the dynamical properties of the networks are directly implied from both the scaling of the strain fluctuations and the mechanical behavior of purely elastic spring networks. 

The mechanical response of spring networks depends sensitively on the   
network's coordination number $z$, the average number of springs 
attached at a node in the network, not including dangling springs. Maxwell's constraint counting argument 
indicates a critical condition of $z_c = 2d$ for the onset of mechanical 
rigidity in a central force network in $d$ dimensions \cite{Maxwell1864}. To 
create a network with variable $z$ in a range spanning 
from well below $z_c$ to well above, springs are arranged on a triangular 
lattice and are removed with a probability $1 - p$ such that the network 
connectivity is roughly $z\simeq 6p$, resulting in a network architecture distinguishable 
from jammed networks~\cite{Ellenbroek2008}. Using units in which the spring rest 
length $\ell_0$ and stiffness $\mu$ are both $1$, the energy can be written 
for small relative deformations ${\bf u}_{ij} = {\bf u}_j - {\bf u}_i$ between neighboring nodes $i$ and $j$ as 
\begin{align}
  \mathcal{H} &=  \frac12  \sum\limits_{\langle ij \rangle}
  g_{ij}({\bf u}_{ij} \cdot {\bf \hat{r}}_{ij})^2
\end{align}
where $g_{ij}=\mu = 1$ for a present bond or $0$ 
for an absent bond and ${\bf \hat{r}}_{ij}$ is a unit vector directed along the 
$ij$-bond in the undeformed reference lattice. This network is embedded in 
a viscous fluid in the low-Reynolds number limit. As the network is deformed, 
hydrodynamic interactions between nodes are ignored, and the fluid  
deforms affinely. Consequently, the net force on a node $i$ is given by the viscous 
Stokes drag and the elastic central forces due to the springs to which it is connected,
\begin{align}
\label{eq:force}
  {\bf f}_i &=4 \pi \eta 
  a (\dot{{\bf u}}_{i} - \dot{{\bf u}}_{\rm fluid}) + \sum\limits_{\langle j \rangle}
  g_{ij} ({\bf u}_{ij} \cdot {\bf \hat{r}}_{ij}) {\bf \hat{r}}_{ij} 
\end{align}
where $\dot{{\bf u}}_{\rm fluid}$ is the velocity field of the underlying 
fluid, $\eta$ is the fluid's viscosity, and the summation of $\langle j \rangle$ is over nearest neighbors of 
node $i$.  For the 2D drag coefficient in the Stokes term, we associated 
a hydrodynamic radius $a$ to a network node and chose natural units such that 
$\eta a=1$. 

\begin{figure}
  \centering
  \includegraphics[width=\columnwidth]{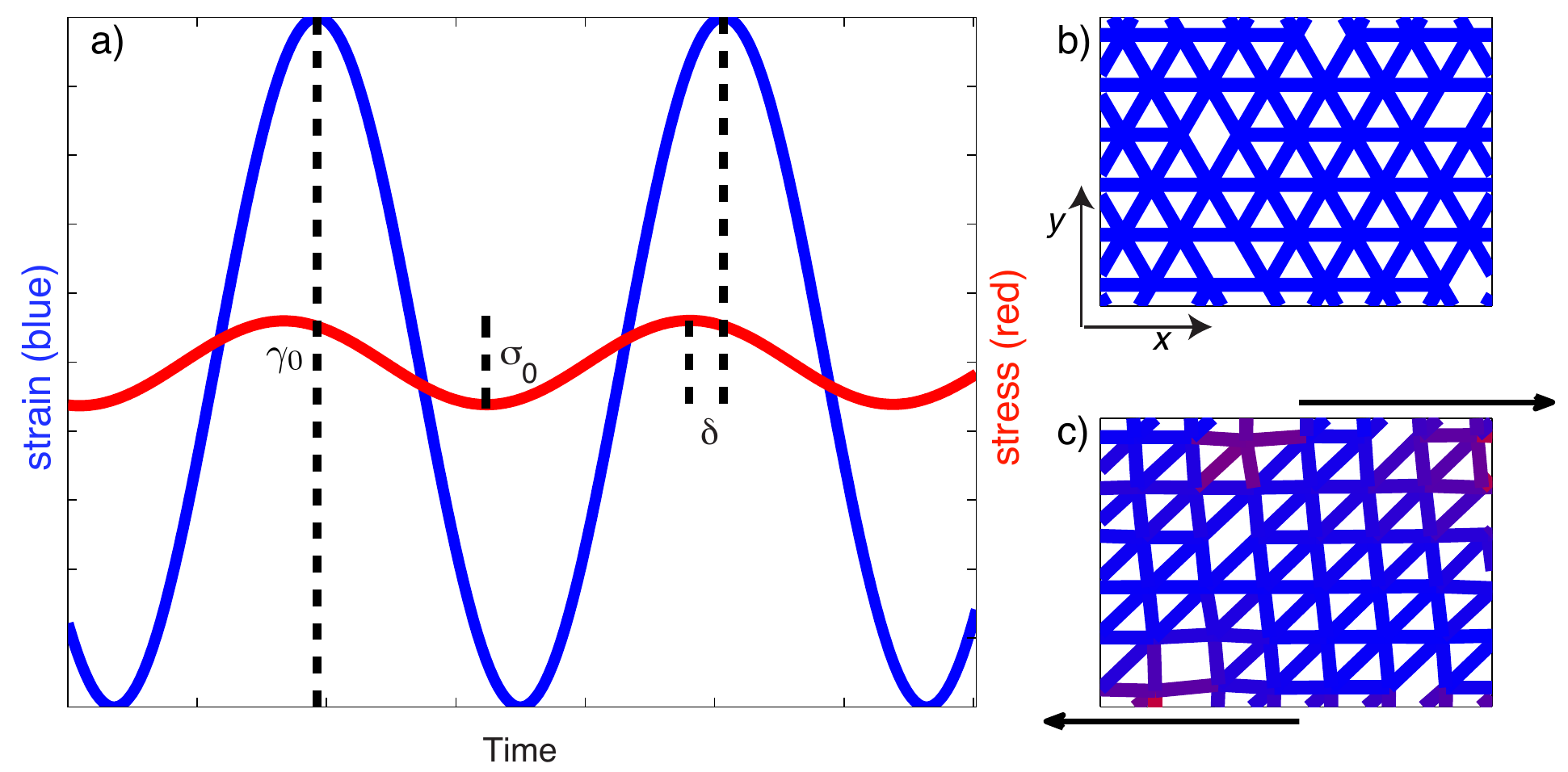}
  \vspace{-0.15in}
  \caption{Plot of a typical strain (blue) applied to the spring network  and
    the resultant stress (red) of the network for $p = 0.6$ (a). We observe
    a phase shift $\delta$ between the applied strain and the response, as
    well as a decrease in amplitude $\sigma_{0}/\gamma_{0}$, from which we can 
    calculate the complex shear modulus given in
    Eq.~\eqref{complexshearmodulus}. The effect of the dynamical shear strain is 
    shown for a sample network on the right; (b) shows the initial network 
    state at $t_0 = 0$, and (c) shows the network state after a quarter 
    oscillation at $t = \pi/2\omega$.
    }
  \label{fig:stress-strain}
\end{figure}

\begin{figure*}
  \begin{center}
    \includegraphics[width=2\columnwidth]{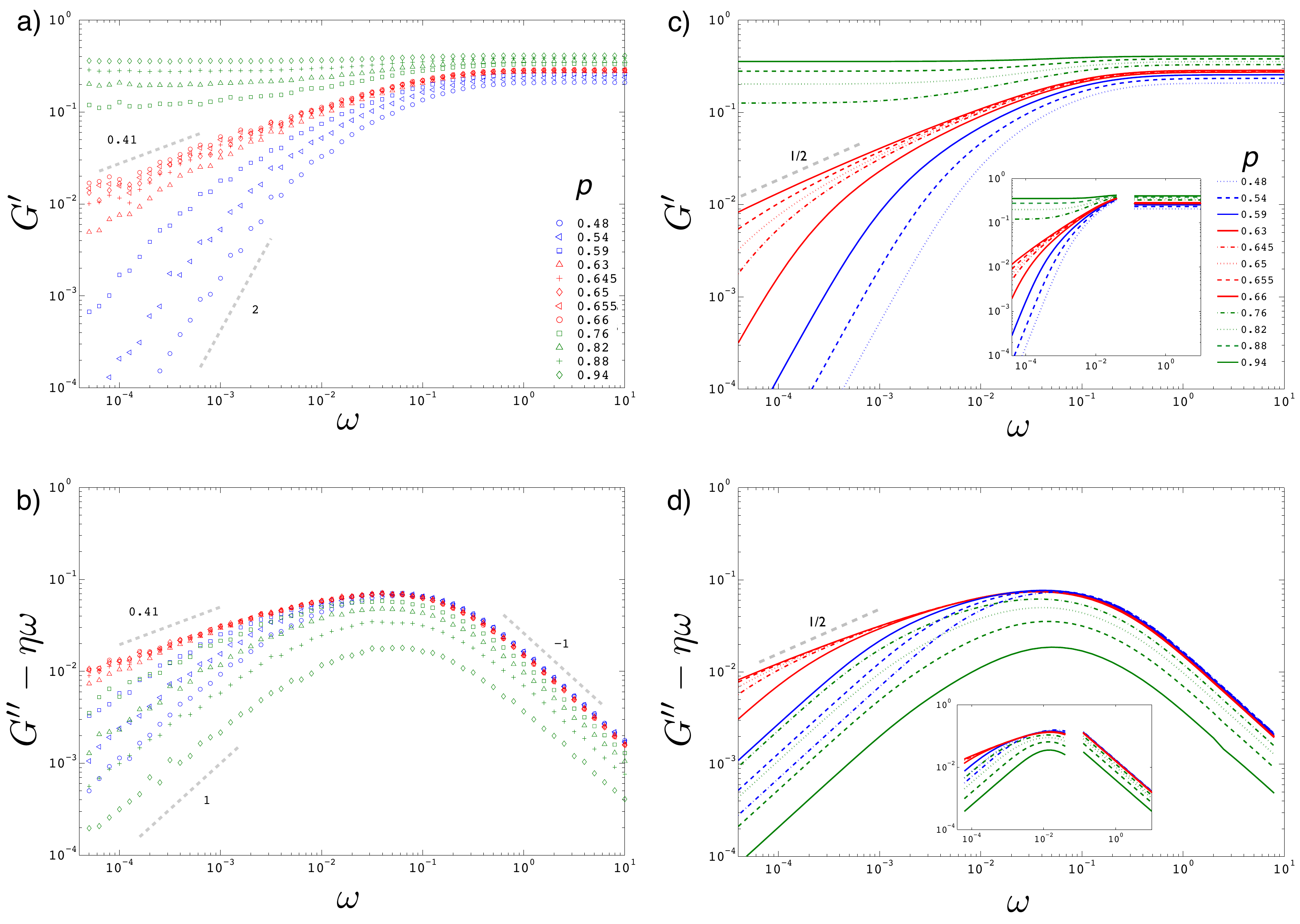}
    \vspace{-0.15in}
    \caption{Simulated rheology for various values of 
           $p$ and $\omega$. All simulations were performed 
       on a network of size $100 \times 100$. The frequency and storage ($G'$) 
       and loss ($G''$) moduli are in natural units, as described in the main text. A slope of $2$ in a) and $1$ in b) indicates low-frequency fluid-like behavior.
       At frequencies below $\omega^*$ near $p_c \approx 0.649$, the shear 
       modulus scales as a power law with exponent $0.41$. Additionally, the EMT results for the storage 
       modulus c) and for the loss modulus d) are shown, in reasonable agreement 
       with the simulations. The insets show the low- and high-frequency limits according to Eqs. \eqref{GemtLow} and \eqref{Gemt}.
    }
    \label{fig:rheodata}
  \end{center}
\end{figure*}

To study the rheology of these diluted networks numerically, we impose 
a time-dependent oscillatory shear strain along the two 
parallel sides of the network with frequency $\omega$ (Fig.~\ref{fig:stress-strain}~bc), which is applied by 
using periodic, Lees-Edwards boundary conditions \cite{Lees1972}. The resulting macroscopic shear stress is 
calculated by 
\begin{equation}
  \label{stresscalc} \sigma_{xy} = \eta \dot{\gamma}(t) + \frac{1}{2 A} \sum\limits_{\langle ij \rangle} 
 {f}_{ij,x} u_{ij,y}
\end{equation}
where $A$ is the surface area of the network and the x-component of the force between two nodes $f_{ij,x} = g_{ij}({\bf u}_{ij} \cdot {\bf \hat{r}}_{ij}) { \hat{r}}_{ij,x}$. A typical stress-strain relationship for an oscillating network
is depicted in Fig.~\ref{fig:stress-strain}~a. From this, we 
determine the complex shear modulus 
\begin{equation}\label{complexshearmodulus} G^*(\omega)= G'( \omega ) + iG''( \omega 
    ) = \frac{\sigma_{0}}{\gamma_{0}} \left[ \cos( \delta (\omega)) + i\sin( \delta (\omega)
        )\right], \end{equation}
where $\sigma_0$ is the magnitude of the 
observed shear stress, $\gamma_0$ that of the imposed shear strain, and 
$\delta (\omega)$ the phase lag between the stress and the strain at frequency $\omega$. Solid-like systems are dominated by $G'$, the 
storage modulus, whereas liquid-like systems are dominated by $G''$, the loss 
modulus.

In the quasistatic limit $\omega \to 0$, the behavior of the network is 
determined by the relation between $z$ and $z_c$ or, equivalently, between $p$ and $p_c$; 
the elastic shear modulus vanishes continuously at the isostatic bond probability $p_c$ as $G'\sim \Delta p^f$, where the rigidity exponent $f \approx 
1.4$~\cite{Arbabi1993,Broedersz2011} and $\Delta p=p-p_c$.  By contrast, in the high 
frequency limit $\omega \rightarrow \infty$, the network's response becomes 
affine, $G \to G_{\rm affine} = p\sqrt{3}/4$, for all bond 
probabilities. 
(Fig.~\ref{fig:rheodata}~a). At high frequencies, nonaffine deformations are 
suppressed by large drag forces between the network and the affinely deforming 
fluid; thus, at high strain rates the fluid effectively dictates the behavior 
of the network. 

\begin{figure}
  \begin{center}
    \includegraphics[width= \columnwidth]{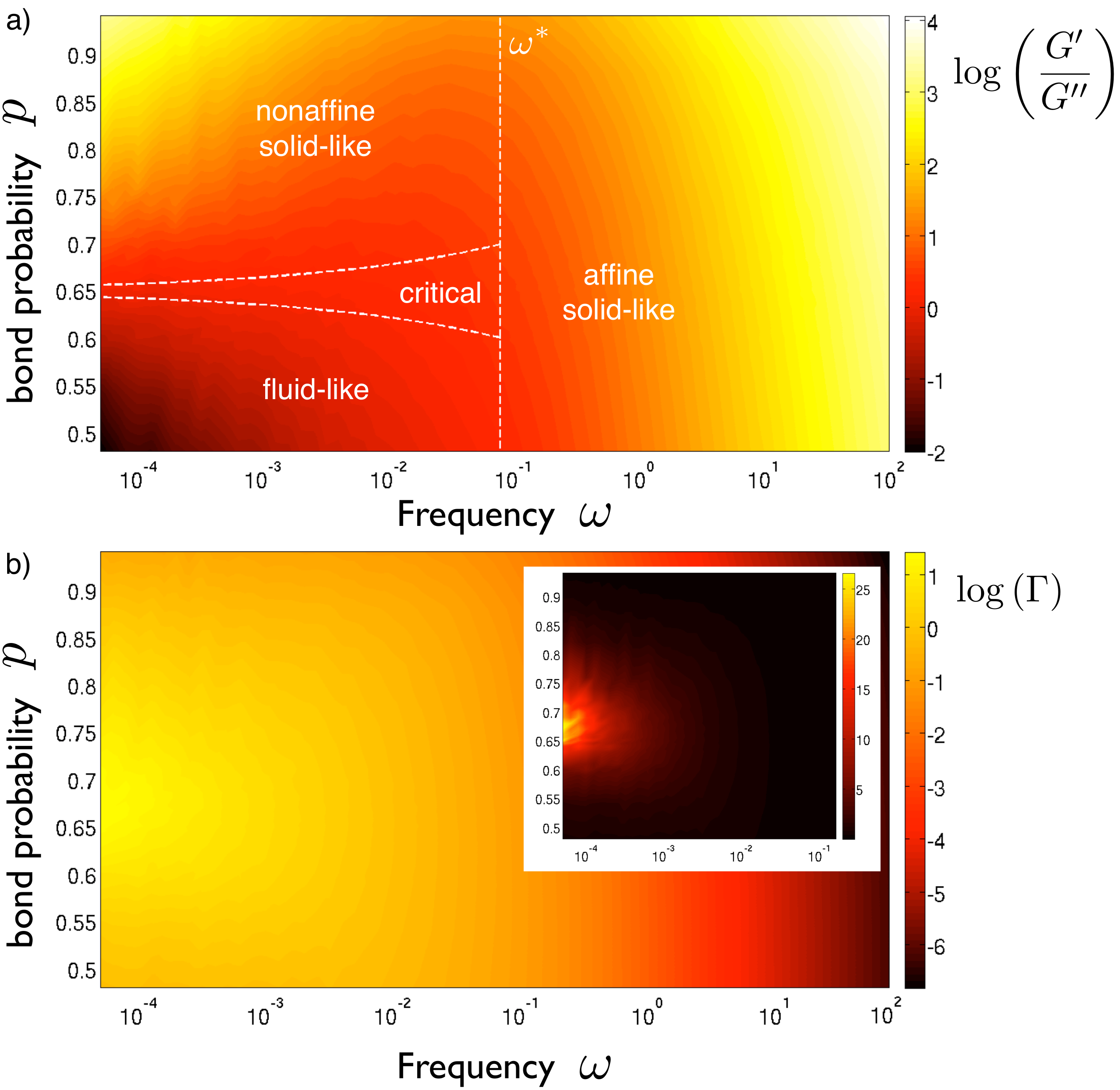}
    \vspace{-0.1in}
    \caption{The dynamical behavior of bond-diluted spring networks. In the high frequency limit, $\omega > \omega^*$, the network 
    deforms affinely for all connectivities; therefore, $G' \gg G''$ (panel~a), and the nonaffinity measure of the network $\Gamma$ vanishes as $\omega^{-2}$ (panel~b).
    Below the crossover frequency $\omega^*$, networks with $p \gg p_c$ are non-affine, solid-like gels with $G' \gg G''$, and 
    networks with $p \ll p_c$ exhibit a more fluid-like response with $G' \ll G''$. In the neighborhood of $p_c$, there is a critical regime where $G' \sim G'' \sim (i\omega)^\Delta$ 
    where $\Delta = 0.41$. The 
    extent of this regime is bound by $\alpha^{-1}_\pm|\Delta p|^{\phi_\omega}$, where $\alpha_\pm$ has been inferred from the collapse described in Fig.~\ref{fig:rheocollapse}.
    Panel~b) shows $\log(\Gamma)$ for various $p$ and $\omega$, illustrating the divergence of the nonaffinity measure at $p_c$ in the low-frequency limit. The inset shows $\Gamma$ on a linear scale.
     }
    \label{fig:heatmap}
  \end{center}
\end{figure}

Networks with connectivities well below the isostatic connectivity behave as Maxwell 
fluids---the storage modulus vanishes as $\omega^2$ and the loss modulus as 
$\omega$ at low frequencies, crossing over at a frequency $\omega^* = 1/4\pi$, set by comparing the stretch modulus of a spring to the drag coefficient of a node, to high-frequency affine elastic behavior. By contrast, hyperstatic 
networks cross over at this characteristic frequency from a nonaffine to an affine solid-like gel at $\omega^*$. 
This transition is accompanied by a maximum in the loss modulus, $G''$ 
(Fig.~\ref{fig:rheodata}~b). At connectivities near isostaticity, the shear 
modulus appears to exhibit a power law regime
$G^*\sim (i \omega)^\Delta$, where $\Delta \approx 
0.41$, extending to the zero-frequency limit as $z \to z_c$. Spring networks in jammed configurations display a similar rheological 
behavior but with mean field exponent $\Delta =1/2$~\cite{Tighe2011,Tighe2012, Tighe2012b}.
The distinct regimes observed here over a connectivity range $3\le z \le 6$ are visualized by plotting the inverse loss tangent $G'/G''$ for a range of network connectives 
and frequencies, as shown in Fig.~\ref{fig:heatmap}~a. 
\begin{figure}
  \begin{center}
    \includegraphics[width= \columnwidth]{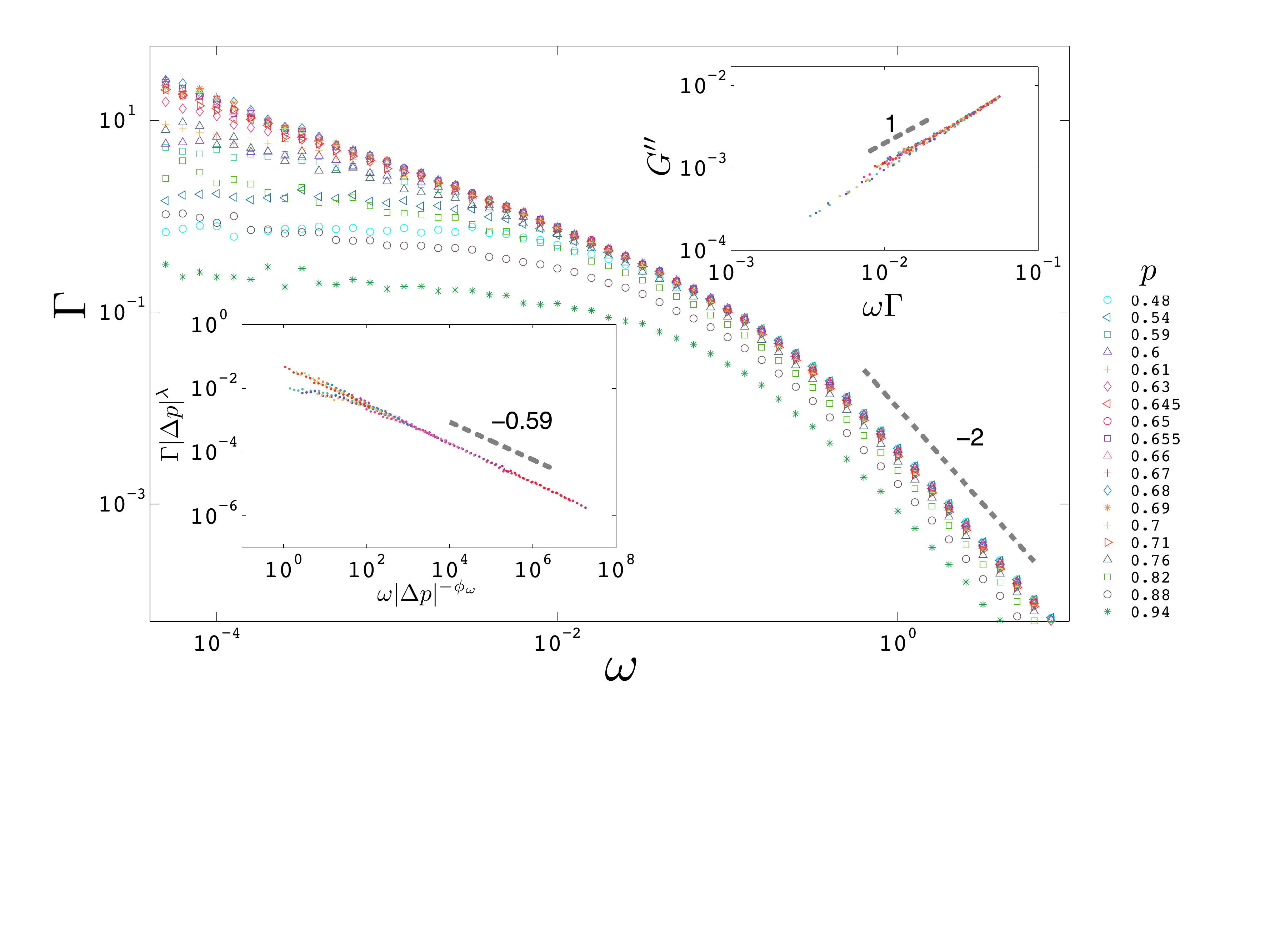}
    \vspace{-0.1in}
    \caption{The nonaffinity measure obtained from simulations as 
        a function of frequency (main plot), collapsed by the scaling 
        form $\Gamma=\left|\Delta p\right|^{-\lambda} \Psi_\pm \left(\omega \left|\Delta p\right|^{-\phi_\omega} \right)$ 
        (lower-left inset), and $G''$ against $\Gamma \omega$ (upper-right 
        inset). At $\omega^* = 1/4\pi$, the nonaffinity crosses over from 
        a high-frequency regime, in which nonaffine fluctuations are 
        minimized by the fluid network with scaling exponent $-2$, to 
        a low-frequency regime, where the nonaffinity measure 
        scales with scaling exponent $-\delta = -0.59$. The extent of 
        this latter region extends to zero frequency as the bond probability approaches the isostatic 
        point. 
    }
      \label{fig:NA}
  \end{center}
\end{figure}

To supplement the complex shear modulus as a description of the macroscopic 
behavior of spring networks, we study the fluctuations in the microscopic network deformations using a simple one-point nonaffinity measure based on the nonaffine component of a node's displacement, $\delta 
{\bf u}$:
\begin{equation}
    \Gamma \equiv \left\langle \frac{\delta {\bf u}^2 }{\gamma^2} 
    \right\rangle,
\end{equation}
where the brackets indicate an average over network nodes and time.
Studies on elastic networks have shown that the nonaffine fluctuations diverge at $p = p_c$ like 
$\Gamma = \Gamma_\pm |\Delta p |^{-\lambda}$~\cite{Wyart2008a,Broedersz2011,Sheinman2012a}, where \emph{$\lambda \approx 2.2$} for bond-diluted 2D triangular networks~\cite{Broedersz2011}. 
In dynamical networks, nonaffine fluctuations will be suppressed by the affinely deforming viscous fluid. Consistent with prior work~\cite{Tighe2012}, our simulations show that the nonaffine fluctuations exhibit a frequency dependence in certain regimes, as shown in Figs.~\ref{fig:NA}~and~\ref{fig:heatmap}~b: far away from $p_c$, the nonaffine fluctuations are frequency-independent at low $\omega$. By contrast, near isostaticity, the nonaffinity measure depends on frequency as a power law $\Gamma\sim \omega^{-\delta}$, with $\delta= 0.59$. For all values of $p$, the nonaffinity vanishes as $\Gamma \sim \omega^{-2}$ beyond the crossover frequency $\omega^*$. The frequency and connectivity dependence of the nonaffinity in dynamic networks can be captured by the scaling ansatz~\cite{Tighe2012}: 
\begin{equation}
\label{eq:naansatz}
\Gamma=\left|\Delta p\right|^{-\lambda} \Psi_\pm \left(\omega / \omega_c \right),
\end{equation}
where the critical relaxation frequency, $\omega_c = |\Delta p|^{\phi_\omega}$, describes the slowest relaxation rate in the system.
Indeed, we find a good collapse with this scaling form using $\lambda=2.2$ as determined previously in elastic networks~\cite{Broedersz2011} and $\phi_\omega=3.6$, as shown in the lower inset of Fig.~\ref{fig:NA}. Near isostaticity, the nonaffine fluctuations are finite, and $\Gamma \sim \omega^{-\delta}$; therefore, it becomes clear that $\Psi_\pm(x) \sim x^{-\delta}$ with $\delta = \lambda/\phi_\omega$ to eliminate the $\Delta p$ dependence in Eq.~\eqref{eq:naansatz}. 

The nonaffine fluctuations can be related to the shear modulus by estimating the dissipated power in the system in the critical regime in two different ways~\cite{Tighe2012,Tighe2012b}. The viscous forces scale as $ \sqrt{\Gamma} \gamma_0 \omega$, giving rise to a dissipated power $W\sim \Gamma (\gamma_0 \omega)^2$, while on a macroscopic level the dissipated power is given by $W= \frac{1}{2} G'' \omega \gamma_0^2$. It follows that $G'' \sim \Gamma \omega$. To test this relation over a broad range of connectivities and frequencies, we plot $G''$ against $\Gamma \omega$ and find that all data collapse onto a curve with a slope of $1$, affirming this correlation~(Fig.~\ref{fig:NA}~upper~inset). Furthermore, since near isostaticity $G'' \sim \omega^\Delta$ with $\Delta \approx 0.41$ (Fig.~\ref{fig:rheodata}~ab), this connection between $G''$ and $\Gamma$ implies that the dynamical exponents $\delta$ and $\Delta$ are related as 
\begin{equation}
\label{eq:delta}
\delta=1-\Delta,
\end{equation}
 which is consistent with our observation of $\delta = 0.59$ (Fig.~\ref{fig:NA}~inset).

\section{Dynamic effective medium theory}

To provide insight into the simulated dynamic rheology, we use an Effective Medium approach~\cite{Feng1985,Schwartz1985,Das2007,Das2012,Sheinman2012a,Mao2011,Mao2013}, a technique dating back to Bruggeman's model for the AC conductivity of disordered composite media~\cite{Bruggeman1935,Clerc1990}. More recently, dynamic effective medium theories have been developed for mechanical networks~\cite{Wyart2010,During2013}. This approach is based on the construction of a mapping from a lattice network where the spring between nodes $i$ and $j$ has a spring constant $g_{ij}$, drawn from a probability distribution $P(g_{ij})$, onto a lattice with uniform, frequency-dependent bond stiffness $\widetilde{g}(\omega)$; this effective lattice mimics the mechanical response of the disordered network at the same global strain, $\epsilon (\omega)$. To derive an expression for $\widetilde{g}\left(\omega\right)$, we follow the procedure in refs~\cite{Wyart2010,During2013}, extending the approach by Feng et al~\cite{Feng1985} by determining the dynamic, effective bond stiffness from a self-consistency requirement, as detailed below.

The effective medium network is subjected to a macroscopic infinitesimal oscillating strain $\epsilon(\omega) = \epsilon_0 e^{i\omega t}$, deforming bond $nm$ by $ {\bf \hat{r}}_{nm} \epsilon(\omega)$. Subsequently, replacing this effective medium bond with one sampled from the distribution $P(g)$ gives rise to an additional, nonaffine deformation $\delta {\bf u}(\omega)$. The original, uniform deformation can be restored by applying a force
\begin{equation}
{\bf f}(\omega) = {\bf \hat{r}}_{nm} \epsilon(\omega)(\widetilde{g}(\omega) - g)
\end{equation}
Thus, the nonaffine deformation which arose from the bond replacement can be expressed as
\begin{equation}
 \delta {\bf u}(\omega) = \frac{{\bf f}(\omega)}{g_{EM}(\omega) - \widetilde{g}(\omega) + g}
\end{equation}
where $g_{EM}\left(\omega\right) $ is the force on a
bond in the effective medium network in response to a unit displacement. This allows us to express the nonaffine displacement as 
\begin{equation}
\delta\mathbf{u}\left(\omega\right)=\frac{\mathbf{\hat{r}}_{nm}\epsilon(\omega)\left(\widetilde{g}\left(\omega\right)-g\right)}{g_{EM}\left(\omega\right)-\widetilde{g}\left(\omega\right)+g},\label{eq:DisplacementMainText}
\end{equation}

The self-consistency condition requires that, when averaging over all possible bond replacements, the local 
fluctuations in the deformation field must vanish, 
$\left\langle \delta\mathbf{u}\left(\omega\right)\right\rangle =0$,
leading to the following equation for $\widetilde{g}(\omega)$, 
\begin{equation}
\int_{0}^{\infty}\frac{g-\widetilde{g}\left(\omega\right)}{g_{EM}(\omega)+g-\widetilde{g}\left(\omega\right)}P\left(g\right)dg=0.\label{eq:mEff_Integral-1}
\end{equation}
We solve this equation by first determining  $g_{EM}^{-1}(\omega)$ as the displacement in response to a unit force between nodes $ n $ and $ m $, $\mathbf{f}\left(\mathbf{k}\right)=\mathbf{\hat{r}}_{nm}\left(1-e^{i\mathbf{k}\cdot\mathbf{\hat{r}}_{nm}}\right)$,  
by solving the network's equation of motion
\begin{equation}
\mathbf{u}\left(\mathbf{k}\right)=-D^{-1}\left(\mathbf{k}\right)\cdot\mathbf{f}\left(\mathbf{k}\right),
\end{equation}
where the dynamical matrix of the effective medium is
given by 
\begin{equation}
D_{nm}=\begin{cases}
-\widetilde{g}\left(\omega\right)\mathbf{r}_{nm}\otimes\mathbf{r}_{nm} & n\neq m\\
\underset{m\neq n}{\sum}\widetilde{g}\left(\omega\right)\mathbf{r}_{nm}\otimes\mathbf{r}_{nm}+4\pi\eta a i \omega\mathbb{I} & n=m
\end{cases},
\end{equation}
 where $\mathbb{I}$ is the unit tensor and $\otimes$ is the external
product. As before, we set $\eta a=1$,  and the spatial Fourier transform of $D$ is given
by 
\begin{eqnarray}
D\left(\mathbf{k}\right) & = & \underset{ij}{\sum}D_{ij}e^{i\mathbf{k}\cdot\mathbf{r}_{ij}}\nonumber \\
 & = & \underset{\mathbf{r}}{\sum}\widetilde{g}(\omega)\mathbf{r}_{ij}\otimes\mathbf{r}_{ij}\left(1-e^{i\mathbf{k}\cdot\mathbf{r}}\right)+i4\pi\omega\mathbb{I}
\end{eqnarray}
Thus, the displacement of the $nm$ bond due to a unit force follows 
\begin{eqnarray}
\label{eq:GEM}
&g&_{EM}^{-1} (\omega) =  \frac{1}{N}\mathbf{r}_{nm}\cdot\underset{\mathbf{k}}{\sum}\mathbf{u}\left(\mathbf{k}\right)\left(e^{-i\mathbf{k}\cdot\mathbf{r}_{nm}}-1\right) = \\
&=&-\frac{1}{N}\underset{\mathbf{k}}{\sum}\mathbf{r}_{nm}\cdot\mathbf{f}\left(\mathbf{k}\right)D^{-1}\left(\mathbf{k}\right)\left(e^{-i\mathbf{k}\cdot\mathbf{r}_{nm}}-1\right) =\nonumber \\
&=&\frac{2\widetilde{g}^{-1}(\omega)}{\mathcal{Z}N} \underset{\mathbf{k}}{\sum}Tr\left[\frac{\underset{\mathbf{r}}{\sum}\mathbf{r}_{ij}\otimes\mathbf{r}_{ij}\left(1-e^{i\mathbf{k}\cdot\mathbf{r}}\right)}{\underset{\mathbf{r}}{\sum}\mathbf{r}_{ij}\otimes\mathbf{r}_{ij}\left(1-e^{i\mathbf{k}\cdot\mathbf{r}}\right)+4\pi\frac{i\omega}{\widetilde{g}(\omega)}\mathbb{I}}\right] =\nonumber\\
&=&\frac{2d}{\mathcal{Z}\widetilde{g}}\left\{1-\frac{i4\pi\omega}{dN\widetilde{g}}\underset{\mathbf{k}}{\sum}Tr\left[ \frac{1}{\underset{\mathbf{r}}{\sum}\mathbf{r}_{ij}\otimes\mathbf{r}_{ij}\left(1-e^{i\mathbf{k}\cdot\mathbf{r}}\right)+4\pi\frac{i\omega}{\widetilde{g}}\mathbb{I}}\right] \right\} \nonumber
\end{eqnarray} 
where $\mathcal{Z}$ is the maximum coordination of the underlying lattice, $ d $ is the dimension of the system and $N$ is the total number of nodes in the network.
 
For a random bond-diluted lattice, the self-consistency condition (Eq.~\eqref{eq:mEff_Integral-1}) can be written as 
\begin{equation}
\label{eq:mEff_diluted}
p\frac{\mu-\widetilde{g}\left(\omega\right)}{g_{EM}(\omega)+\mu-\widetilde{g}\left(\omega\right)}-\left(1-p\right)\frac{\widetilde{g}\left(\omega\right)}{g_{EM}(\omega)-\widetilde{g}\left(\omega\right)}=0,
\end{equation} 
where $\mu$ will be set to $1$. 
By solving this equation for a triangular lattice configuration, we obtain the macroscopic shear modulus, $G^* (\omega)= \widetilde{g}(\omega) \sqrt{3}/4$. Remarkably, this EMT prediction for the rheology captures the main  features of the simulation results with reasonable quantitative agreement, as shown in Figs.~\ref{fig:rheodata}cd.
Slow convergence of the numerical integration scheme precludes a high-precision solution of the EMT in the critical regime. However, we can obtain various interesting analytical results by considering  the large or small limits of  the quantity $\left|\widetilde{g}(\omega)\right|/\omega$.

\paragraph{High-frequency limit}
When $ \omega \gg \left|\widetilde{g}(\omega)\right|$, Eq.~\eqref{eq:GEM} can be written as $g_{EM} \approx 2\pi \omega i$. Using this in the self-consistency equation, we find the shear modulus to be 
 \begin{equation}
G^* (\omega)\approx \frac{\sqrt{3} p}{4}\left(1+i\frac{1-p}{2 \pi  \omega}\right)
\label{GemtLow}
\end{equation}
This high frequency limit corresponds quantitatively with the numerical results, as shown in the insets of Figs.~\ref{fig:rheodata}cd.

\paragraph{Low-frequency limit}
This limit is solvable only when $ p $ is not much less than $ p_c $. In this case $\left|\widetilde{g}(\omega)\right|\gg \omega$ and Eq.~\eqref{eq:GEM} reduces to
\begin{equation}
g_{EM}^{-1}(\omega) \approx \frac{2\widetilde{g}^{-1}(\omega)}{3} \left[1-\frac{2\pi i \omega}{\widetilde{g}(\omega)}\mathcal{A}\right]
\end{equation}
where $ \mathcal{A} $ is a numerical constant, 
\begin{equation}
 \mathcal{A} = \frac{1}{N}\underset{\mathbf{k}}{\sum}Tr\frac{1}{\underset{\mathbf{r}}{\sum}\mathbf{r}_{ij}\otimes\mathbf{r}_{ij}\left(1-e^{i\mathbf{k}\cdot\mathbf{r}}\right)} \simeq 5.17.
\end{equation} 
By solving the self-consistency equation for a bond-diluted lattice (Eq. \eqref{eq:mEff_diluted}) in this limit, we find the shear modulus, 
\begin{eqnarray}
\label{Gemt}
G^*(\omega)\approx & \frac{\sqrt{3}}{16} \bigg[&6p-4-8 i \mathcal{A} \pi  \omega \nonumber \\
    & & \left. +\sqrt{64 i \mathcal{A} \pi  \omega+\left(6p-4-8 i \mathcal{A} \pi  \omega\right)^2}\right]
\end{eqnarray}
This is consistent with results found by D\"uring et al.~\cite{During2013}. This expression for the dynamic shear modulus captures the  low-frequency rheology for $p\gtrsim p_c$, as shown in the insets of Figs.~\ref{fig:rheodata}~cd.

\begin{figure}
  \begin{center}
    \includegraphics[width= \columnwidth]{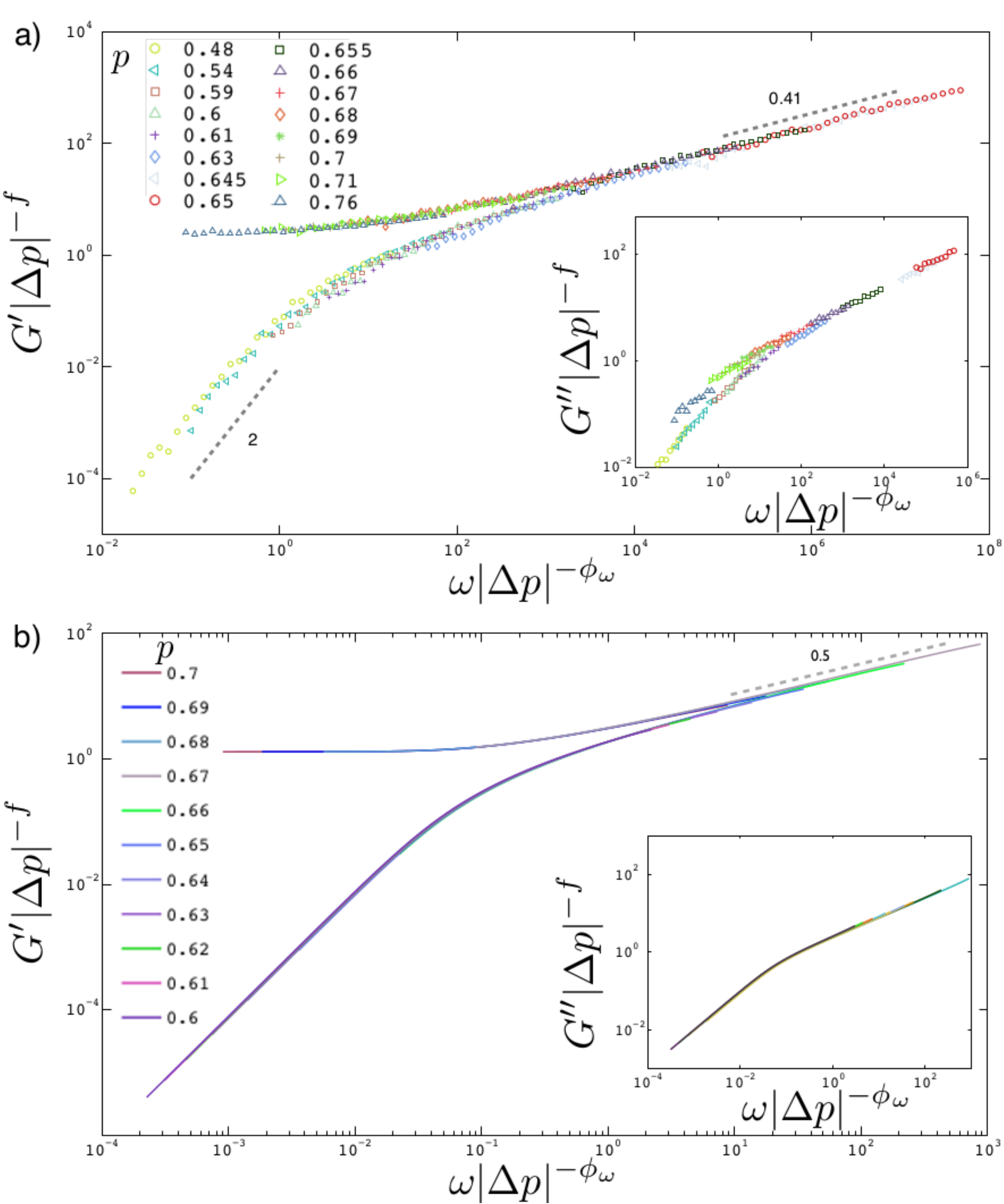}
    \vspace{-0.1in}
    \caption{Universal scaling of the shear modulus for the numerical results (a) and the 
        analytical calculations (b) according to the scaling form $G^* 
        = |\Delta p|^f \mathcal{G}_\pm(\omega|\Delta p|^{-\phi_\omega})$ over 
        a range of bond probabilities. The main plots show the results for the elastic modulus, the insets for the loss modulus.
        The scaling exponents for the EMT calculation are $f = 1$ and 
        $\phi_{\omega} = 2$. An excellent collapse is observed 
        for the numerical data with scaling exponents $f = 1.4$ and 
        $\phi_\omega = 3.6$ and $p_c = 0.649$ for a $100\times 100$ network~\cite{Broedersz2011}. A frequency range of 
        frequencies between $3.1\times 10^{-5}$ and $4.0\times 10^{-2}$ was used for the numerics, and in the EMT scaling the frequency range is $10^{-6}$ to $ \left(2 \mathcal{A} \pi\right)^{-1}$ for $G'$ and $10^{-6}$ to $\left(2 \mathcal{A} \pi\right)^{-2}$ for $G''$.
    }
      \label{fig:rheocollapse}
  \end{center}
\end{figure}

The EMT indicates a critical bond probability at $p_c=2/3$. Close to the critical point, $\left| \Delta p \right| \ll 1 $, and in the limit of small frequencies ($ \omega \ll \left(\mathcal{A} \pi\right)^{-1}\approx 10^{-1}$ for $G'$ and $ \omega \ll \left(\mathcal{A} \pi\right)^{-2}\approx 10^{-2}$ for $G''$), the shear modulus in Eq.~\eqref{Gemt} can be written in the following scaling form~\cite{Tighe2011}
\begin{equation}
\label{eq:widom}
G^*=\left| \Delta p \right|^{f} \mathcal{G}^*_{\pm}\left(\frac{\omega}{\left| \Delta p \right|^{\phi_\omega}}\right)
\end{equation}
where the EMT predicts $f = 1$ and $\phi_\omega=2$, consistent with the mean field predictions in ref.~\cite{Tighe2011, Tighe2012b}. The scaling function  $\mathcal{G}^*_{\pm}\left(x\right)=\mathcal{G}'_{\pm}\left(x\right)+i\mathcal{G}''_{\pm}\left(x\right)$  is given by
\begin{eqnarray}
    \label{eq:scalingfunc}
\mathcal{G}'_{\pm}\left(x\right)&=& \frac{3\sqrt{3}}{8}\left\{ \cos\left[\frac{1}{2} \tan^{-1}\left( \alpha x\right) \right]\left[1+\left( \alpha x\right)^2 \right]^{1/4} \pm 1\right\} \nonumber \\
\mathcal{G}''_{\pm}\left(x\right)&=& \frac{3\sqrt{3}}{8} \sin\left[\frac{1}{2} \tan^{-1}\left( \alpha x\right) \right]\left[1+\left(  \alpha x\right)^2 \right]^{1/4} 
\end{eqnarray} 
with $\alpha=\frac{16 \mathcal{A} \pi}{9}$. The scaling function $ \mathcal{G}'_{\pm} $ has a $+$ branch above $p_c$ and a $-$ branch below $p_c$. When $x\rightarrow 0$ and $p>p_c$ $\mathcal{G'_+}(x)$ must be constant such that $G'$ scales as $|\Delta p|^f$, while $\mathcal{G''_\pm}(x) \sim x$ so that $G''$ scales as $\omega |\Delta p|^{f - \phi_\omega}$. Furthermore, when $x\rightarrow 0$ and $p<p_c$ $\mathcal{G'_-}(x) \sim x^2$ such that $G'$ scales as $\omega |\Delta p|^{f-2\phi_\omega}$, and $\mathcal{G''_-}(x) \sim x$ such that $G''$ scales as $\omega |\Delta p|^{f - \phi_\omega}$. 
In the critical connectivity regime, the shear modulus is finite and, thus, $\mathcal{G^*_\pm}\left( x\right) \sim x^{\frac{f}{\phi_\omega}}$ such that $G^*$ is $ \Delta p $-independent. Consequently, $G^*\sim(i \omega)^\Delta$ with $ \Delta=f / \phi_\omega $. 

The EMT and numerical results are collapsed according to this scaling form, as shown in Fig.~\ref{fig:rheocollapse}. We find an excellent collapse for the numerical data with $f=1.4$, determined in prior work on elastic networks~\cite{Arbabi1993,Broedersz2011}, and $\phi_\omega=3.6$, determined from collapsing the nonaffinity data (Fig.~\ref{fig:NA}). Within the EMT, $\Delta=1/2$~\cite{Wyart2010,During2013}, consistent with the mean field calculations in refs.~\cite{Tighe2012, Tighe2012b}. By contrast, from the collapse of the simulation data, we find $\Delta\approx 0.41$. This difference between the exponents predicted by the EMT and our numerical results are due to the mean-field nature of the effective medium approximation. Specifically, the effective medium theory assumes small nonaffine fluctuations \cite{During2013}. This assumption appears to be justified for most network connectivities and frequencies, as shown by the good comparison between the EMT and numerical results shown in Fig.~\ref{fig:rheodata}. However, the nonaffine fluctuations become large as the network approaches criticality; in the quasistatic limit, such fluctuations are expected to diverge for networks with dimension greater than or equal to $2$ \cite{During2013}. Therefore, any approximation that does not take these diverging fluctuations into account cannot be expected to predict the correct exponents.

The scaling behavior of $G^*$ is clearly related to that of the nonaffinity parameter $\Gamma$: in both cases, the critical relaxation frequency is controlled by the exponent $\phi_\omega$. In the first case, we inferred that $\phi_\omega = f/\Delta$, while in the second case, we found that $\phi_\omega = \lambda / \delta$, and thus, 
\begin{equation}
\frac{f}{\Delta} = \frac{\lambda}{\delta}
\end{equation}
Solving for $\Delta$ and recalling that $\delta = 1 - \Delta$ (Eq.~\ref{eq:delta}), we obtain
\begin{equation}
\label{eq:epic}
\Delta = \frac{f}{\lambda + f}
\end{equation}
Strikingly, we find that the dynamical behavior of the network, captured by the exponent $\Delta$, can in fact be inferred from the rigidity exponent $f$ and the nonaffinity exponent $\lambda$ of the elastic network in the absence of viscous interactions. Using previously determined $f = 1.4 \pm 0.1$ and $\lambda = 2.2 \pm 0.4$~\cite{Broedersz2011}, we expect $\Delta = 0.39 \pm 0.08$, in agreement with our numerical observations. Furthermore, using Eq.~\eqref{eq:epic}, we can also recover the mean-field  prediction for the nonaffinity exponent $\lambda = 1$, using $\Delta = 1/2$ and $f = 1$ from the EMT calculation~\cite{Wyart2008a}. This argument should be valid for broader classes of disordered networks, e.g.\ fiber networks for which $f_b = 3.2\pm0.4$ and $\lambda_b = 1.8 \pm 0.3$ in 2D, implying a dynamical scaling of $\Delta_b = 0.64 \pm 0.13$. These results show that viscous interactions act like a field taking the system away from criticality. Furthermore, similar scaling arguments have been constructed, relating the dynamic conductivity of disordered resistor networks to the exponents that govern the DC response~\cite{Efros76}. Thus, the dynamic EMT, combined with the scaling arguments, provide an avenue for exploring dynamical behavior of a wide range of disordered networks.

\begin{acknowledgments}
We thank Cliff Brangwynne for helpful discussions and his hospitality and Brian Tighe, Fred MacKintosh and Boris Shklovskii for insightful discussions. 
This work was supported by the Lewis-Sigler fellowship (CPB) and FOM/NWO (MS). 
\end{acknowledgments}

\end{document}